\documentclass[useAMS,usenatbib,12pt]{mn2e} 
\usepackage{graphicx}
\usepackage{txfonts}

\def\mh1{$M_{\rm H_{I}}$}

\title[H{\sc i} in the interacting galaxy Arp86]{H{\sc i} content and star formation in the interacting galaxy Arp86}
\author [Sengupta {\it{et al.}}]{ Chandreyee Sengupta $^{1}$\thanks{e-mail:sengupta@ncra.tifr.res.in(CS),
 dwaraka@rri.res.in(KSD), djs@ncra.tifr.res.in(DJS)}, K. S. Dwarakanath$^{2}$ and D. J. Saikia$^{1}$ \\\\
$^{1}$National Centre for Radio Astrophysics -Tata Institute of Fundamental Research, Pune 411 007, India\\
$^{2}$Raman Research Institute, Bangalore 560 080, India}

\begin{document}

\date{}
\pagerange{\pageref{firstpage}--\pageref{lastpage}} \pubyear{}

\maketitle

\label{firstpage}

\begin{abstract}
We present the results of Giant Metrewave Radio Telescope (GMRT) observations 
of the interacting system Arp86 in both neutral atomic hydrogen, H{\sc i}, and in radio 
continuum at 240, 606 and 1394 MHz. In addition to H{\sc i} emission from the 
two dominant galaxies, NGC7752 and NGC7753, these observations show a complex
distribution of H{\sc i} tails and bridges due to tidal interactions. The regions
of highest column density appear related to the recent sites of intense star
formation. H{\sc i} column densities $\sim$1$-$1.5 $\times$10$^{21}$ cm$^{-2}$ have been detected in the tidal bridge 
 which is bright in Spitzer image as well. We also detect H{\sc i} emission from the galaxy  2MASX J23470758+2926531, 
which is shown to be a part of this system. We discuss the possibility that this could 
be a tidal dwarf galaxy. The radio continuum observations show evidence of a non-thermal
bridge between NGC7752 and NGC7753, and a radio source in the nuclear region
of NGC7753 consistent with it having a LINER nucleus. 
\end{abstract}

\begin{keywords}
galaxies:interactions - galaxies:dwarf - galaxies:spiral - galaxies:Arp86 - radio lines:galaxies - radio continuum:galaxies
\end{keywords}

\begin{table*} 
\caption{Some of the optical properties of Arp86 }
\begin{tabular}{llr}
\hline
&NGC7753 &NGC7752\\
\hline
Major Diameter$^{a}$&3.3$^{\prime}$ &0.8$^{\prime}$ \\
Minor Diameter$^{a}$&2.1 $^{\prime}$ &0.5$^{\prime}$ \\
Classification$^{a}$&SAB(rs)bc  &I0 \\
Radial Velocity$^{b}$&5160 kms$^{-1}$ & 4940 kms$^{-1}$\\
Distance&68 Mpc  &68 Mpc \\
K band mag$^{c}$ (absolute)&$-$25.5 &$-$22.9 \\ 
B band mag$^{c}$ (absolute)&$-$22.1  &$-$19.7 \\

 \hline
$a$:Nasa Extragalactic Database\\
$b$: \cite{marcelin}\\
$c$:Apparent magnitude values from Nasa Extragalactic Database
\end{tabular}
\end{table*}

\begin{table*} 
\caption{GMRT observations }
\begin{tabular}{lllllllllr}
\hline
 Frequency & Observation  & Phase      & Phase cal    &  $\tau$    & Bandwidth &rms (per channel  & beam size   \\ 
           & date         & calibrator &  flux density        &           &           &for 21 cm line)   &      \\
           &              &            & (Jy)    & (hrs)    & (MHz)         &      (mJy beam$^{-1}$)  &  (arcsec $\times$ arcsec)               \\     
\hline

 21 cm line & 2008 May 01  &J0029+349 &2.0 & 9  & 8&  0.6 &11 $\times$ 11  \\
                                            &  &  & & & & 0.8 &25 $\times$ 25  \\
                                            &  &  & & & & 1.0 &40 $\times$ 40  \\
  1394 MHz& 2008 May 01 & J0029+349 &2.0 &- &- &0.4  &16 $\times$ 16  \\
  606 MHz  & 2008 May 23 &J0137+331  &29.5&3 &16 & 0.6 & 16 $\times$ 16\\
  240 MHz  &2008 May 23 & J0137+331  & 51.8&3 & 6& 1.5 &16 $\times$ 16 \\
\hline

\end{tabular}
\end{table*}

\section {Introduction}

Galaxy interactions and mergers have been known to affect galaxy evolution in various ways and one such aspect is star formation in a galaxy. Since the early seventies, several theoretical and observational studies have been conducted to understand how star formation in a galaxy is altered or affected by collisions and mergers \citep{larson,biermann,woods,overzier}. Several such studies have shown that prolonged enhancement of star formation is the cause of much of the infrared emission in major mergers \citep{sanders,miho,lin}. Minor mergers or tidal interactions, which are common in moderate density environments like galaxy groups, also seem to affect the star formation rates and morphologies of galaxies. Many of these systems have been detected to have tidal bridges, tails and debris containing large amounts of H{\sc i} with blue optical counterparts. However, it was not always clear whether the blue colour of these optical counterparts were due to the star forming disk material pulled out in the interaction or due to in situ star formation in the tidal debris. The advent of ultraviolet (UV) and mid infrared (MIR) telescopes like Galex and Spitzer have revealed bright recent star forming clumps in the tidal debris \citep{Smith, Hibbard, Neff}.  
      H{\sc i} observations of such systems are important as these give us an estimate of the gas densities necessary to trigger star formation. Recent observations show evidence of star formation in remote sites away from the main disk, like tidal bridges and debris. The Magellanic bridge has been observed to host star formation in regions with N(H{\sc i}) $\sim$10$^{20}-$ 10$^{21}$ cm$^{-2}$ \citep{Muller, Harris}. The Arp's loop in 
M81$-$M82 system is undergoing recent star formation \citep{de Mello}. H{\sc ii} regions have been discovered in several systems, embedded in the tidal features \citep{Ryan-Weber}. Studying in situ star formation in tidal debris is of importance as this can also enrich the intergalactic medium (IGM) in addition to the galactic wind 
scenario \citep{Ryan-Weber}. \cite{Ryan-Weber} estimate that star-formation rate (SFR) as low as 1.5$\times$10$^{-3}$ M$_{\odot}$ yr$^{-1}$ maintained for 1 Gyr, can pollute the IGM with a metallicity $\sim$1$\times$10$^{-3}$ solar. This value compares well with the ``metallicity floor" $\sim$1.4$\times$10$^{-3}$ solar in the damped Lyman alpha (DLA) gas, observed over a redshift range of 0.5 to 5 \citep{Prochaska}.  
  
In order to study the H{\sc i} gas properties such as the morphology, kinematics and column 
density distributions, and their correlation with the star forming zones, especially in the tidal 
bridges, tails and debris, we have presently chosen the remarkable but rather less studied 
interacting system Arp86. It is an archetypal example of interacting galaxies consisting of a 
grand-design barred spiral galaxy, NGC7753, with a small companion, NGC7752, towards the end
of one of the spiral arms, much like the extensively studied M51 system. NGC7753 has been classifed as
SAB(rs)bc in the NASA Extragalactic Database (NED) and as SB(r)bc by \cite{Franco-Balderas}, while NGC7752 is classified as I0 by
NED and as Irr galaxy by \cite{Nilson}.  The optical diameter of NGC7753 is 3.3$^{\prime}$ and that of NGC7752 is
0.8$^{\prime}$.  \cite{marcelin} have determined the heliocentric velocities
of NGC7753 and NGC7752 to be 5160 and 4940 km s$^{-1}$ respectively, while the values listed in NED
are 5168 and 5072 km s$^{-1}$ respectively. 
The distance to the system, using their average optical velocity and a Hubble constant of 75 km s$^{-1}$ Mpc$ ^{-1}$ 
is 68 Mpc (1$^{\prime\prime}$ $\sim$0.33 kpc). 
\cite{keel} note that in their sample of spiral galaxies,
NGC7753 is unusual in the small extent of the nuclear emission which appears unresolved. They classify
NGC7753  as a low-ionization nuclear emission region (LINER), while the brightest H{\sc ii} complex
which contains many H$\alpha$ knots occurs towards NGC7752. \cite{Smith} present Spitzer MIR
images of this system which is a part of their sample of interacting systems to study interaction-induced
star formation. The Spitzer MIR observations show the presence of active star forming regions 
in the spiral arms of NGC7753, extending all the way to NGC7752, in the form of a tidal bridge. Optical 
images also show the tidal bridge connecting the two galaxies, with bright regions of star formation in
the bridge (Laurikainen, Salo \& Aparicio, 1993). Some of the optical properties of Arp86, are summarised in Table 1.  

In order to investigate the correlation of gas properties and star formation in interacting systems,
we observed this interesting system with the Giant Metrewave Radio Telescope (GMRT) in both H{\sc i}
and radio continuum at 240 and 606 MHz. The observations are described in Section 2, the observational
results are presented in Section 3, while the results are discussed in Section 4. The main results
are summarised in Section 5.   

\begin{figure}
     \hbox{
    \centerline{\rotatebox{0}{\includegraphics*[height=4in]{A86-2DY-LILR-40ASECMOM0-edited.PS}} }
    \hspace{-8.2cm}
    \centerline{\rotatebox{-90}{\includegraphics*[height=3.0in]{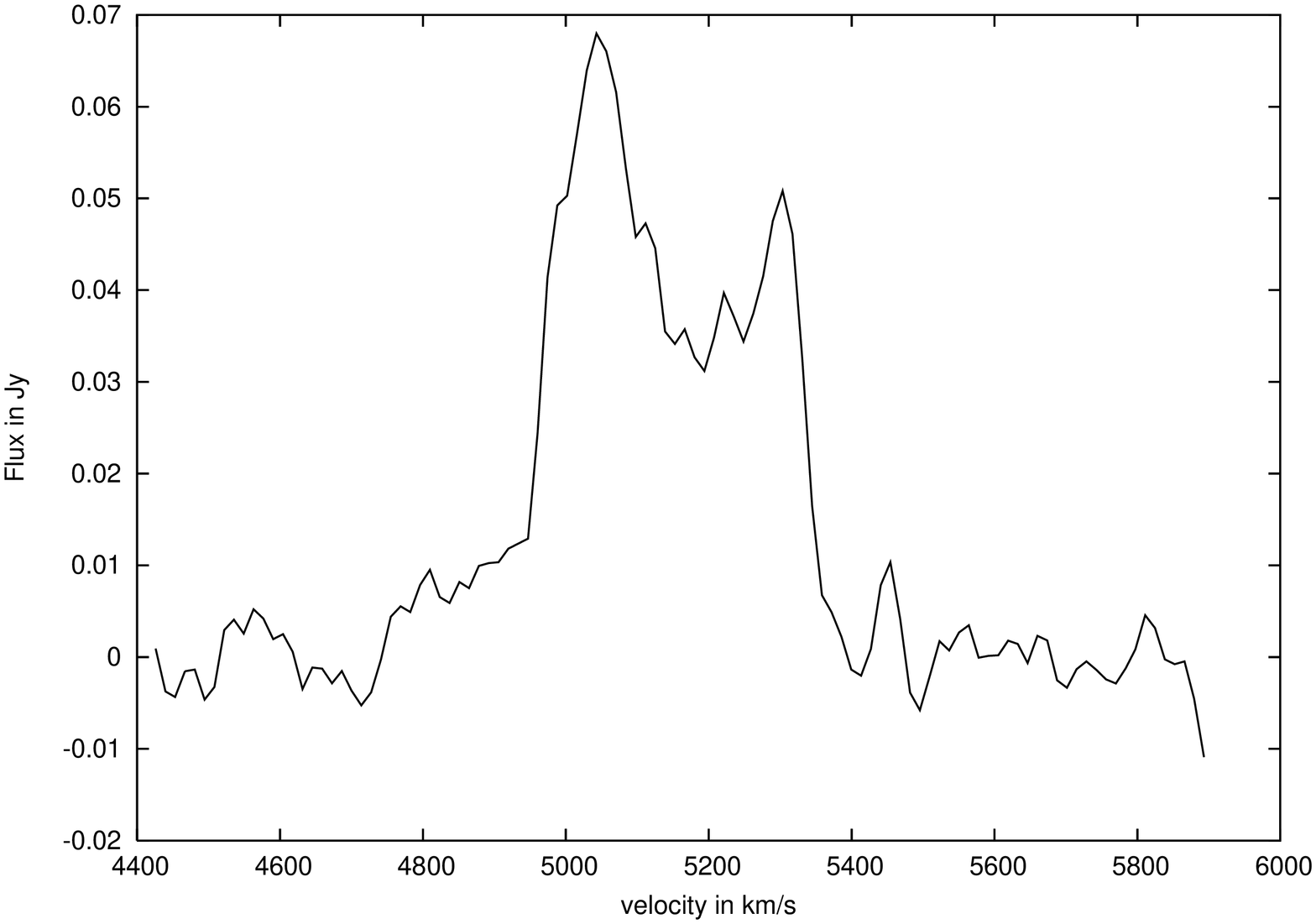}} }
          } 
    \caption{Upper pannel : H{\sc i} column density map of Arp86 with an angular resolution of 40$^{\prime\prime}$ overlayed on an optical Digitized Sky
     Survey (DSS) map. The H{\sc i} column density contours are 1.4$\times10^{19} \times$ (3, 7, 15, 25, 40, 50, 70) cm$^{-2}$.
    Lower pannel : The GMRT H{\sc i} spectrum of Arp86 obtained with an angular resolution of 40$^{\prime\prime}$.}
    \label{}
    \end{figure}

\begin{figure*}
    \hspace{.1cm}
    \centerline{\rotatebox{0}{\includegraphics*[height=8.0in]{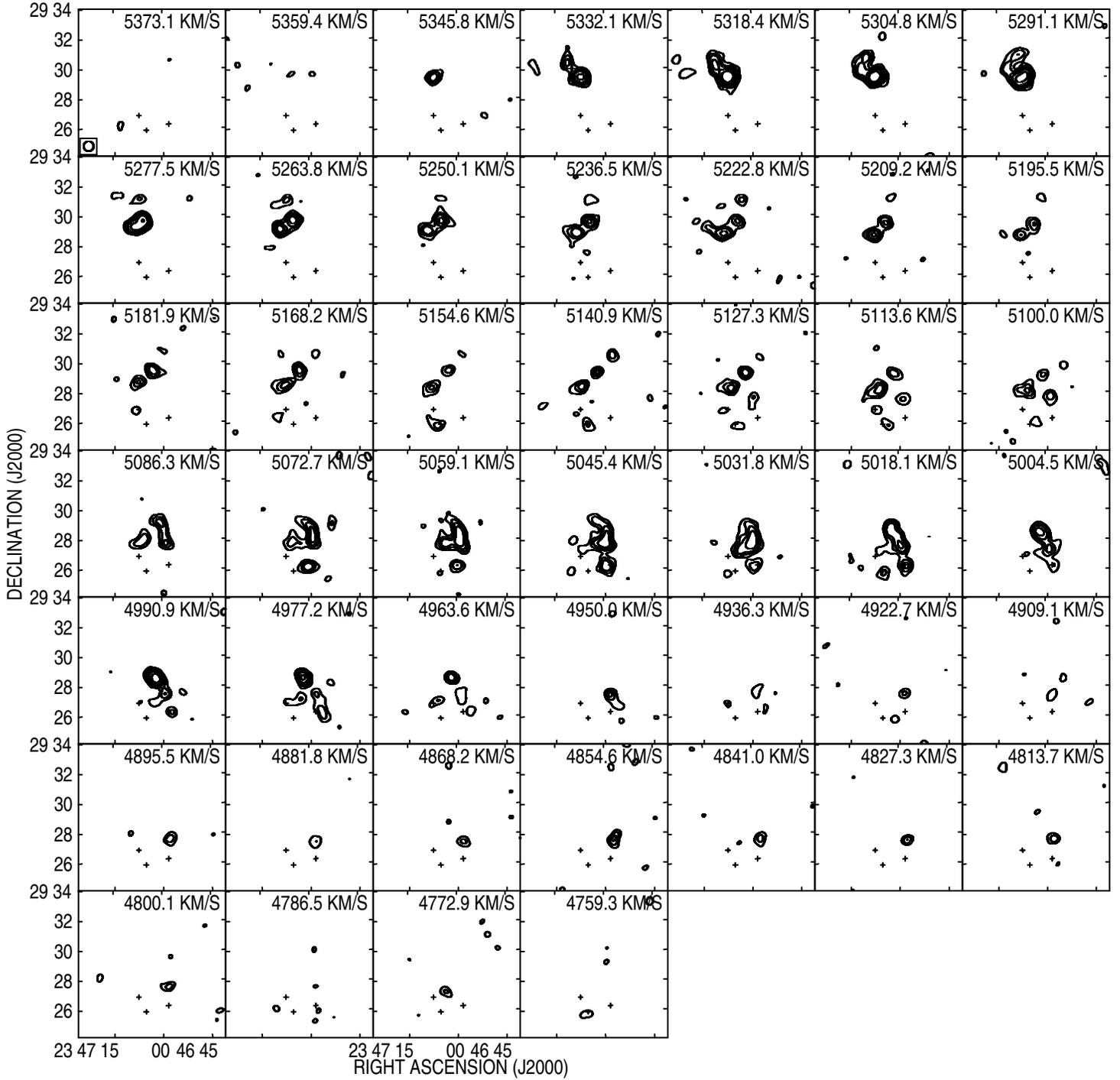}} }
    \vspace*{0.05in}
    \caption{The channel images of Arp86 with an angular resolution of 40$^{\prime\prime}$. The contour levels are 1 mJy/beam $\times$(3,5,7,10,15). The first plus sign (from left) denotes position of 2MASX J23470758+2926531, the next two `+' signs denote the two H{\sc i} peaks in the south eastern H{\sc i} extension. The signs help to understand the location of H{\sc i} in this region.}     \label{}
    \end{figure*}

\section {Observations}

Arp86 was observed for 9 hours in H{\sc i} 21 cm line and for 3 hours each at 606 and 240 MHz, in cycle 14 of GMRT observations. 
The GMRT is an interferometric array of 30 antennas, each of 45-m diameter, spread over a maximum baseline of 25 km. At frequencies of $\sim$1420 MHz, the system temperature and the gain (K/Jy) of the instrument are 76K and 0.22 respectively. System temperatures for 610 MHz and 235 MHz are 102K and 237K and gains for these two frequencies are 0.32 and 0.33 respectively. The full width at half maximum of the primary beam of GMRT antennas is $\sim$24\arcmin ~at 1420 MHz,  43\arcmin ~ at 610 MHz and 114\arcmin~ at 235 MHz. The baseband bandwidth used was 
8 MHz for the 21cm H{\sc i} line observations (velocity resolution $\sim$13.7 km s$^{-1}$), 16 MHz for the 610 observations and 6 MHz for the 235 MHz observations. The observing log and the observational details are summarized in Table 2, which is
self explanatory. The pointing centre for all
the observations was 23$^{\rm h}$ 47$^{\rm m}$ 01.$^{\rm s}$61 +29$^\circ$ 28$^\prime$ 17.0$^{\prime\prime}$  in J2000 co-ordinates. The
observations were done in the standard way with the phase calibrator observed before and after each scan on the source. The primary
flux density and bandpass calibrator was 3C286 and 3C48 with an estimated flux density in the standard Baars flux density scale \citep{baar}.

Data obtained with the GMRT were reduced using {\tt AIPS} (Astronomical Image Processing System). 
Bad data due to dead antennas and those with significantly lower gain than others, and radio frequency interference (RFI) 
were flagged and the data were calibrated for amplitude and phase using the primary and secondary calibrators. 
The calibrated data were used to make both the H{\sc i} line images and the 20 cm radio continuum images 
by averaging the line-free channels and self calibrating. 
For the H{\sc i} line images the calibrated data were continuum subtracted using the AIPS tasks `UVSUB' and `UVLIN'. 
The task `IMAGR' was then used to get the final 3-dimensional deconvolved H{\sc i} data cubes. From these cubes the 
total H{\sc i} images and the H{\sc i} velocity fields were extracted using the AIPS task `MOMNT'.
While reducing the data at 240 and 606 MHz, a similar procedure as described above was followed. To avoid bandwidth smearing, 
the available baseband bandwidth was divided into smaller parts and used as input to the imaging task `IMAGR'. 
Multiple facets were used to cover the primary beam at both the lower frequencies. For both the frequencies, 
the field was divided into 31 facets and imaged. To bring out the structures on different levels in both H{\sc i} and
radio continuum, we produced images of different resolutions by tapering the data to different uv limits.  

\section {Observational results}

\subsection {H{\sc i} morphology}
Fig. 1 presents the total H{\sc i} column density map of the system  with our lowest resolution of 
40$^{\prime\prime}$, overlayed on DSS optical map and the integrated spectrum from the same data cube. 
A synthesised beam of this size samples the system with a spatial resolution of $\sim$13 kpc. The galaxy to the north east is 
NGC7753 and that to the south west is NGCC7752. 
The total H{\sc i} map reveals a very disturbed morphology, the presence of a 
tidal bridge between the two galaxies and H{\sc i} in form of tidal tails and debris around the system. 
The disk of NGC7753 is rich in H{\sc i}, though the centre shows depletion in the H{\sc i} column density, possibly due to ongoing intense star formation. The bulk of the 
gas follows an arc like structure starting from the north-western edge of the disk of NGC7753, following the optical bridge to 
the companion galaxy NGC7752. Apart from these, there are two prominent spiral-arm like features, one which points initially
towards the north-east and then turns towards the west, while the other one is south of NGC7752 and points towards the east. 
In neither of these two extensive H{\sc i} features, any optical or MIR counterparts are seen. As we will refer to the observations and model of \cite{Laurikainen} and \cite{Salo}, henceforth we will refer to the 
southern extension of H{\sc i} from NGC7753 as the H{\sc i} bridge and the northern extension of H{\sc i} as the tail, 
a nomenclature used in \cite{Laurikainen}. We refer to the feature towards the south of NGC7752, which was not seen in earlier
observations, as the `south-eastern extension'.  Embedded in the 
tidal debris to the east, we detect a small galaxy, listed as 
`2MASX J23470758+2926531' in NED, which did not have any 
previous spectroscopic data. Our observations find this galaxy to be a 
part of this system.

From the spectrum shown in Fig. 1 which has a velocity resolution of 13.7 km s$^{-1}$,  
we estimate an integrated flux density of 18.7 Jy km s$^{-1}$ which is smaller than the value of 
22.7 Jy km s$^{-1}$ estimated from  single-dish observations \citep{huct}, suggesting that there could be more diffuse emission
which is not visible in our image. We detect 2.1 $\times$ 10$^{10}$ M$_{\odot}$ of H{\sc i} mass from the Arp86 system 
using our lowest-resolution image.

\begin{figure}
    \centerline{\rotatebox{0}{\includegraphics*[height=4in]{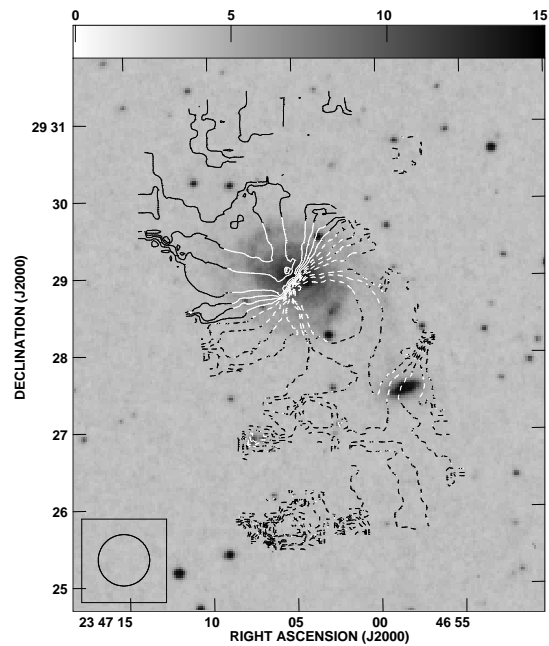}} }
    \caption{H{\sc i} velocity field of Arp86 with an angular resolution of 40$^{\prime\prime}$, overlayed on an optial DSS image. 
     Velocity levels : ($-$300, $-$250, $-$200, $-$160, $-$130, $-$110, $-$70, $-$50, $-$30, $-$10, 10, 30, 50, 70, 110, 130, 140) km s$^{-1}$ 
     relative to the central velocity of 5168 km s$^{-1}$.}
    \label{}
    \end{figure}

The channel maps and the first-moment image, which represents the intensity-weighted velocity field, 
with an angular resolution of 40$^{\prime\prime}$ are shown in Figs. 2 and 3 respectively. The H{\sc i} channel maps show a 
clear connection between NGC7753 and NGC7752 in velocity space. The spiral feature in the north of NGC7753, referred to as the tail,  is visible at 
velocities ranging from $\sim$5330 km s$^{-1}$ to 5070 km s$^{-1}$, with the highest velocities occuring towards the
north-eastern end of NGC7753. The gas in the disk of this galaxy shows signs of rotation, with the south-eastern side 
approaching us. The gas in the bridge connecting the two galaxies is approaching us, with the velocities in NGC7752 
extending up to $\sim$4800 km s$^{-1}$.    
In the eastern side, gas is seen towards 2MASX J23470758+2926531 over a range of velocities extending from 4963 to 5181 km s$^{-1}$.  The velocity field  of the
south-eastern extension ranges from $\sim$4936 km s$^{-1}$ south of NGC7752 to $\sim$5150 km s$^{-1}$. This is close to the velocity
seen in the galaxy 2MASX J23470758+2926531. It is interesting to note that there appears to be weak emission seen in the low-resolution
H{\sc i} image (Fig. 1) connecting the edge of the south-eastern feature to the 2MASX galaxy. The nature of this galaxy is
discussed later in the paper.

\begin{figure}
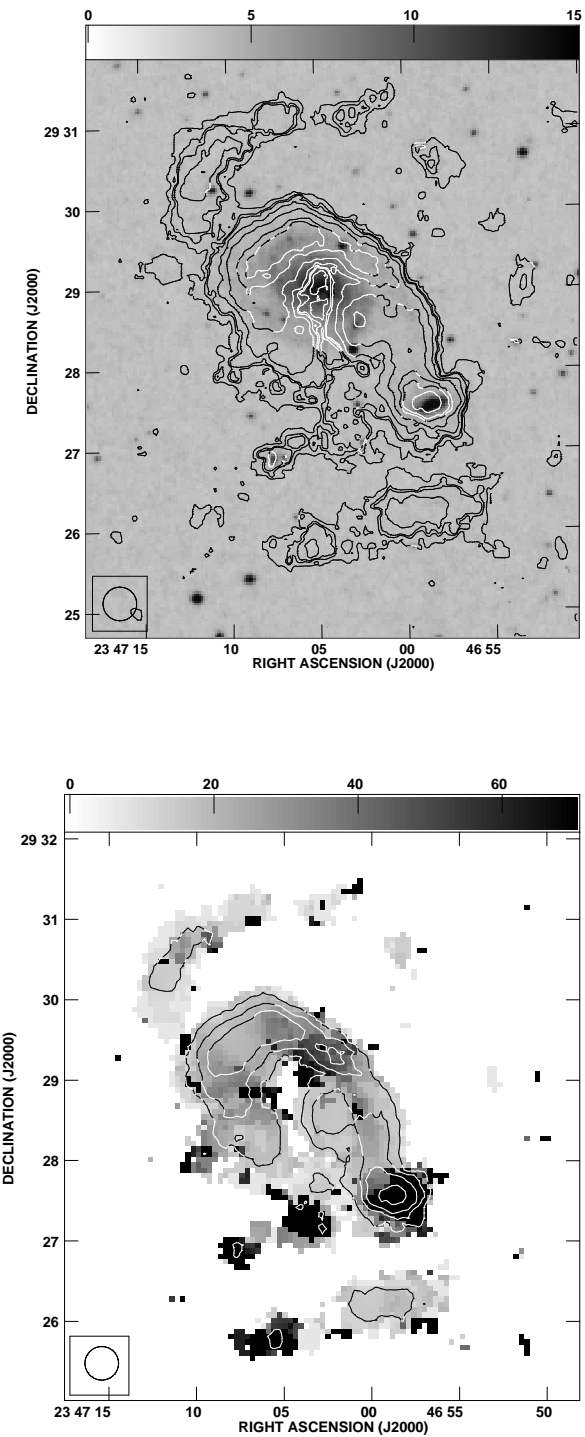

    \hspace{.1cm}
    \centerline{\rotatebox{0}{\includegraphics*[height=4in]{A86-FIG1-TOTALHI-25ASEC-1-edited.PS}} }
    \centerline{\rotatebox{0}{\includegraphics*[height=4in]{A86-FIG5-TOTALHI-DISP-25ASEC-1-edited.PS}} }
    \caption{Upper panel: H{\sc i} column density map of Arp86 with an angular resolution of 25$^{\prime\prime}$ overlayed on an optical Digitized Sky
     Survey (DSS) map. The H{\sc i} column density contours are 2.5$\times10^{19} \times$ (3, 7, 9, 15, 25, 40, 50, 70) cm$^{-2}$.
     Lower panel:  H{\sc i} contours overlayed on grey scale H{\sc i} velocity dispersion map with an angular resolution of 25$^{\prime\prime}$.
     The velocity dispersion range is from 0 to 70 km s$^{-1}$. H{\sc i} column density contours are 1.8$\times10^{19} \times$ (20, 40, 60, 100) cm$^{-2}$.}
    \label{}
    \end{figure}

The H{\sc i} column density image superposed on the optical DSS image, and the velocity dispersion with an angular resolution of
25 arcsec, which corresponds to about 8 kpc, are shown in Fig. 4. The regions with high velocity dispersions of $>$50 km s$^{-1}$
are seen towards NGC7752,  the galaxy    2MASX J23470758+2926531 and towards the eastern end of the `south-eastern extension'. 
There also appears to be a region of high velocity dispersion between NGC7752 and 2MASX J23470758+2926531 but this could be
due to low signal to noise ratio and needs confirmation. The rest of the emission has velocity dispersions less than this value. It may be relevant to note here that in this image
the `south-eastern' extension appears disjointed from the Arp86 system.

To explore possible correlations of star forming regions of the system with H{\sc i} 
column density, we have made an H{\sc i} column density image with a higher angular
resolution of 11 arcsec which corresponds to $\sim$3.6 kpc. The higher resolution helps
minimise dilution of the estimate of the column density from more diffuse emission. 
This high-resolution image is shown superimposed on the 24 micron Spitzer MIR in Fig. 5.
The star-forming regions in the system are seen as the dark patches in the greyscale 
MIR image. The H{\sc i} image shows clumps and knots of emission with peak column
densities in the range of 1$-$4 $\times10^{21}$ cm$^{-2}$, and absence of emission towards the
centre of NGC7753. The regions of high star formation towards the eastern and northern 
edges of NGC7753 and then following the bridge between the two galaxies and also the 
emission from NGC7752 appear to correlate with the regions of highest column density.
 There is also a region of star formation towards the south-western edge of NGC7753,
where high-column density H{\sc i} gas is seen.

\begin{figure}
    \hspace{.1cm}
    \centerline{\rotatebox{0}{\includegraphics*[height=3.8in]{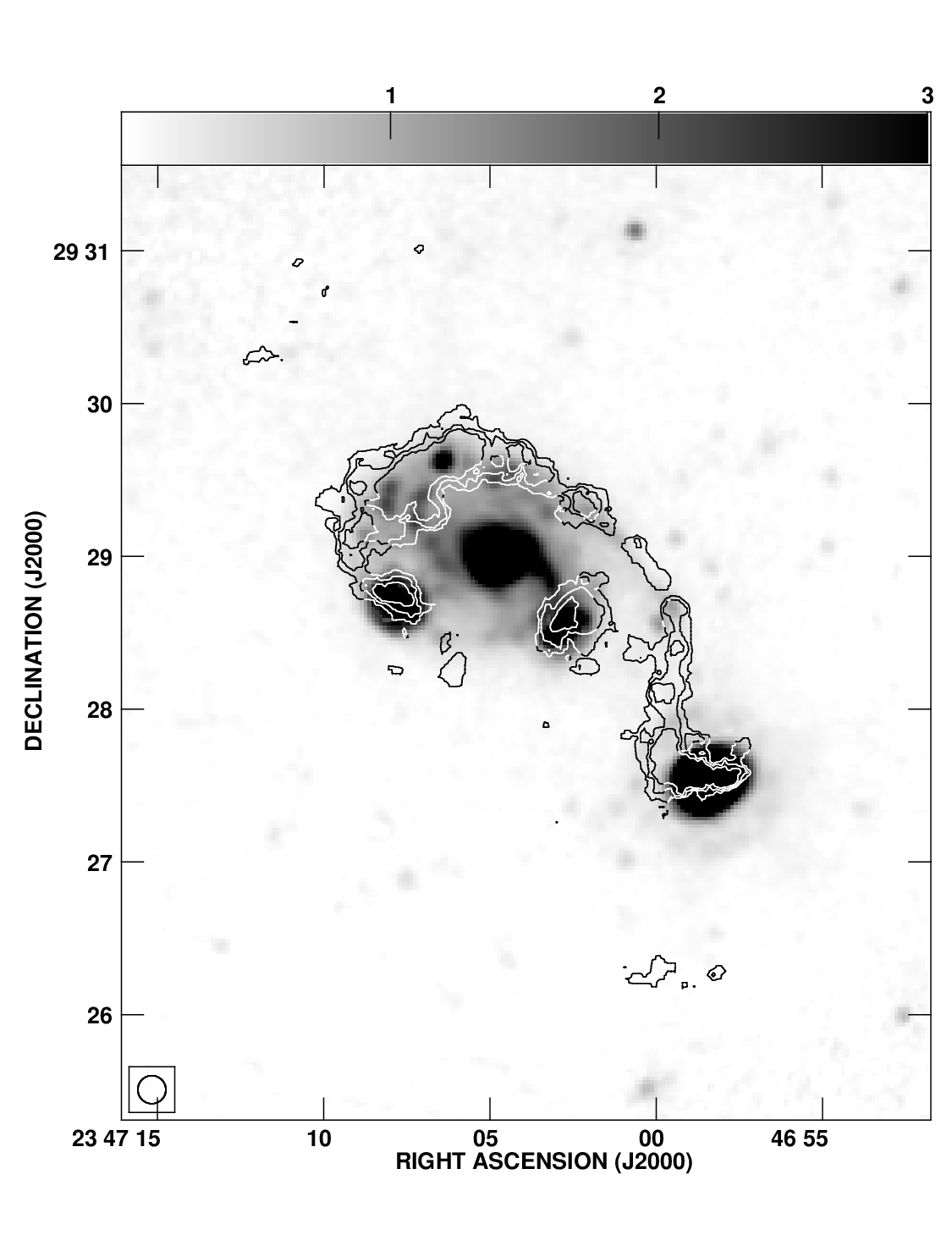}} }
    \vspace*{0.05in}
    \caption{The H{\sc i} column density image of Arp86 with an angular
resolution of 11 arcsec overlayed on the Spitzer 24-micron image.  The H{\sc i}
 column density contours are 1$\times10^{20} \times$(10, 15, 20) cm$^{-2}$.}
    \label{}
    \end{figure}

\begin{figure*}
\hbox{
\includegraphics[scale=0.31,angle=0]{A86-235-45ASEC-edited.PS} 
\includegraphics[scale=0.29,angle=0]{ARP86-610-45ARCSEC-NEW-edited.PS}
\includegraphics[scale=0.29,angle=0]{1420-CONTINUUM-45ARCSEC-edited.PS}
     }
\caption{The radio continumm images of Arp86 with an angular resolution
of 45 arcsec at 240 (left panel), 606 (middle panel) and 1394 (right panel) MHz.}
\end{figure*}

\begin{figure*}
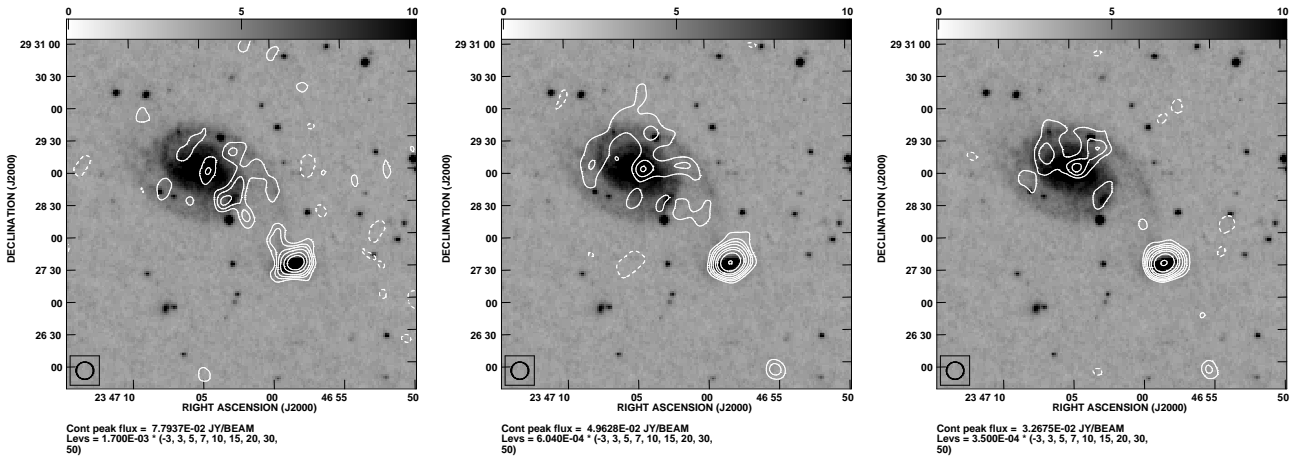

\hbox{
\includegraphics[scale=0.3,angle=0]{A86-235-16ASEC-edited.PS}
\includegraphics[scale=0.3,angle=0]{A86-610-16ASEC-edited.PS}
\includegraphics[scale=0.3,angle=0]{A86-1420-16ASEC-edited.PS}
     }
\caption{The radio continumm images of Arp86 with an angular resolution
of 16 arcsec at 240 (left panel), 606 (middle panel) and 1394 (right panel) MHz.}
\end{figure*}

\subsection {Radio Continuum}

The radio continuum images at 240, 606 and 1394 MHz of NGC7752 and NGC7753 with angular resolutions of 45
and 16 arcsec, which correspond to $\sim$15 and 5 kpc respectively, are presented 
in Figs. 6 and 7 respectively. Amongst the low-resolution 
images, the 606 MHz one which has an rms noise of 0.64 mJy beam$^{-1}$, shows clear 
evidence of a radio continuum bridge connecting the galaxies NGC7752 and NGC7753. 
Although there may be some evidence of this bridge at 240 MHz, it is not seen clearly 
at 1394 MHz (Fig. 6). The higher-resolution images of the system presented in 
Fig. 7 have all been smoothed to the resolution obtained at 240 MHz.
The small galaxy, 2MASX J23470758+2926531, was not detected in radio continuum, in any of 
these three bands.                              

Radio continuum emission from NGC7752 seems to be starformation dominated and shows it to be well resolved at the highest resolution (7$\times$3 arcsec$^2$) image.  The total flux densities obtained from the images
at different frequencies are summarized in Table 3. 

The low-resolution images of NGC7753 show that it is dominated largely by 
diffuse disk emission, although its structure appears somewhat different at the 
different frequencies, which may be at least partly due to different mixtures
of thermal and non-thermal emission. The total flux densities estimated from
the low-resolution images over similar areas are presented in Table 3. The typical 
uncertainties in the flux density estimates at 1394 and 606 MHz are $\sim$5\% and at 
240 MHz $\sim$18\%.  
However, given the quality of the image and the relatively lower surface brightness of the diffuse 
emission from NGC7753, the uncertainty in its flux density could be somewhat higher.
At 240 MHz, the disk emission is seen to have two peaks, both of which correlate 
with the star forming regions in the disk. 
The set of high-resolution maps also show fragmented emission, largely coinciding 
with the star forming regions. These observations also reveal the presence of a 
weak compact central source in NGC7753, which has not been noted earlier. The 
position of this compact source whose
peak at 23$^{\rm h}$ 47$^{\rm m}$ 04.$^{\rm s}$40 +29$^\circ$ 29$^\prime$ 01.2$^{\prime\prime}$ is coincident within the errors with
the optical centre of the galaxy located at 23$^{\rm h}$ 47$^{\rm m}$ 04.$^{\rm s}$8 +29$^\circ$ 29$^\prime$ 00.4$^{\prime\prime}$ (NED). 
The flux densities of the central component estimated from the images in Fig.7
are listed in Table 3. The flux densities of this feature are similar even when
one considers the highest resolution images at 606 and 1394 MHz. The spectra obtained from these measurements are presented in Fig. 8 and are discussed in section 4.6.

\begin{table} 
\caption{Estimates of radio continuum integrated flux densities }
\begin{tabular}{llllllr}
\hline
 Frequency &  beam size  &  NGC7752  & NGC7753    & NGC7753 (core)   \\ 
   MHz & ($^{\prime\prime}$)     & (mJy)          &   (mJy)         & (mJy)           \\
   
\hline

 1393.8 &45 &26 &21 &$-$ \\
        &16 &25  & $-$& 2.6  \\

 606.5  &45 &46 &26 &$-$ \\
        &16 &45  & $-$& 4.5 \\

 240.0  &45 &69 &$\sim$50 & $-$\\
        &16 &76 &$-$ &9.1 \\
  
\hline

\end{tabular}

\end{table}

\section {Discussion}

Arp86 is similar to an M51 type system, with a grand-design main galaxy interacting 
with a small galaxy at the tip of one of its arms. Arp86 has been studied in detail and 
modelled to understand the nature of interaction induced star formation in the system 
\citep{Laurikainen, Salo}. Recent Spitzer observations in the MIR provide quite a 
detailed picture of the tidally induced star formation sites in the system and reveal a 
tidal bridge with ongoing star formation \citep{Smith}. To search for interaction-enhanced star formation in the MIR, instead of comparing luminosities, which measure the absolute SFR, the authors suggest that it is better to compare the Spitzer colours, which measure mass-normalized SFRs. The bands in which these observations were carried out were 3.6, 4.5, 5.8, 8.0 and 24 micron. The 4.5$-$5.8, 3.6$-$8.0, 8.0$-$24, and 3.6$-$24 colours are all measures of the mass-normalized SFR, with a redder colour indicating a higher normalized SFR. The 3.6 and 4.5 micron bands are dominated by the older stellar population, while the other bands have significant contributions from interstellar dust heated by young stars. For the set of M51-like systems in this sample of Arp galaxies, of which Arp86 is a member, the authors note redder 8.0$-$24, 3.6$-$24 and 5.8$-$24 colours and a general enhancement of SFR, especially enhancements that are localised in some regions of the system. This is in agreement with the results of \cite{Laurikainen}, who found bluer optical colors in the central regions of 9 out of 13 M51-like galaxies they studied including a detailed study of Arp86.

\subsection{H{\sc i} morphology}

H{\sc i} observations of interacting systems are very useful to probe
a number of aspects. They enable us to probe the
distribution of gas due to the interactions and constrain theoretical models,
probe regions of star fomation which may be induced by the interactions and examine 
correlations of star formation sites with H{\sc i} column density, and explore the 
formation and properties of tidal dwarf galaxies.

The H{\sc i} features in Arp86 resemble those of M51 in many ways, which
have been modelled as being due to tidal interactions between the two galaxies
\citep{Howard}. As disussed earlier, the low-resolution H{\sc i} images of Arp86 reveal large ($\sim$4\arcmin) gaseous tidal arms and debris, towards both the north and south of the system. An H{\sc i} bridge connects NGC7753 to its smaller mass companion NGC7752. 
H{\sc i} debris seen beyond that, towards the south, may either be an extension of a tidal arm or may be material pulled out from the system and dumped in the IGM. In the channel maps, the southern debris seems to have a continuity from 4936 km s$^{-1}$ to 5154 km s$^{-1}$. A detailed high-resolution H$\alpha$ velocity field study of Arp86 \citep{marcelin}, found a steep velocity gradient in NGC7752 and an anomalous velocity distibution between 4920 km s$^{-1}$ to 5060 km s$^{-1}$ towards its north-eastern region. More importantly the 
north-eastern edge of NGC7752 and the bridge end, where it joins NGC7752 share similar radial velocities  $\sim$5060 km s$^{-1}$. The authors suggested this could be a signature of 
a tidal tail/bridge about to be torn from NGC7752. Our H{\sc i} map reveals tidal debris at this position, matching the velocities suggested by \cite{marcelin}. Though we lack the 
necessary resolution to study the velocity field of NGC7752 in detail, our H{\sc i} channel maps show the southern debris to start at $\sim$4936 km s$^{-1}$, close to the velocity of 
4920 km s$^{-1}$ suggested by \cite{marcelin}. In the lowest-resolution H{\sc i} map (40$^{\prime\prime}$), the southern debris seems to be loosely connected to the main system of Arp86. However, in the 25$^{\prime\prime}$ map, the debris looks detached from the system, 
quite as suggested in \cite{marcelin}. However, even though it appears detached, the debris has a continuity in the velocity space till $\sim$5150 km s$^{-1}$.

Towards the north of Arp86, an H{\sc i} tidal tail of $\sim$4\arcmin ~in length, 
which corresponds to $\sim$80 kpc is seen. It initially points towards the north-east,  
then curves 
towards west and later runs almost parallel to the tidal bridge between NGC7753 and NGC7752.
This feature is continuous in velocity space between 5330 to 5070 km s$^{-1}$. A similar 
feature was noticed in the H{\sc i} images of the M51 system by \cite{Rots}. A long 
H{\sc i} tail without any optical counterpart was seen to be connected loosely to the disk of NGC5194. A detailed simulation involving 3 components, stars, gas clouds and dark halo and taking into account collision of the gas clouds, could reproduce most of the H{\sc i} features of M51 successfully, but not the extended H{\sc i} tail \citep{Howard}. The 
H{\sc i} tail was proposed to be a remnant of the previous passage of the companion. The ratio of the masses of the companion NGC5195 to the main galaxy NGC5194 in case of the M51 
system, needed to be roughly 0.1 for the simulation to be able to reproduce the observational features \citep{Howard}. From the H$\alpha$ rotation curves, the masses of NGC7753 and NGC7752 were estimated to be 1.3$\times$10$^{11}$M$_{\odot}$ and 1.8$\times$10$^{10}$M$_{\odot}$ respectively \citep{marcelin}. This makes the mass ratio of the companion to the main galaxy $\sim$0.1, a value similar to that of the 
M51 system \citep{Howard}. The results from the simulations and rotation curve analysis of \cite{Salo} suggests M51-like interaction in the Arp86 system involving multiple passages
 of the companion around the main galaxy. Perturbation by a companion at closed orbit involving mutiple passages may lead the main disk to exhibit open spiral structures \citep{Salo}. M51 is an example of this, and our observations show this for Arp86 as well. 
Simulation results of \cite{Salo} indicate a spiral arm towards the north of Arp86, but our 
H{\sc i} observations show a much more extended feature than expected in \cite{Salo}. However, taking the example of M51 \citep{Rots,Howard} and taking into consideration 
that the interactions of M51 and Arp86 are of similar nature, the long northern tidal tail of Arp86 can be a remnant from the past passage of the companion.

The velocity field of Arp86 has been studied in detail using H$\alpha$ observations by \cite{marcelin}. Isovelocity contours reveal normal differential rotation in both the disks. However, signature of warping was found in the western side of NGC7753, the position angle of the major axis was seen to bend towards NGC7752. Even NGC7752 was reported to appear warped towards its northeastern edge, where isovelocity contours were found to be quite disturbed, following typical signs of interaction. Our H{\sc i} observations of coarser resolution, show that the  velocity field exhibits regular disk rotation in NGC7753,
with the velocity reaching a maxima at $\sim$5340 km s$^{-1}$ towards the north-east
which is the receding side. These numbers are in agreement with the H$\alpha$ velocity field observed by \cite{marcelin}. On the western side of the NGC7753, the velocity field is disturbed and chaotic possibly due to the interactions. 

\subsection{Star formation in the tidal bridge and tail}

The optical BVRI photometry \citep{Laurikainen} and the Spitzer images 
\citep{Smith} suggest that this system has undergone a recent enhancement of 
star formation, possibly induced by the interaction. \cite{Laurikainen} find the 
bridge to be bluer than the tail. Our high-resolution image of H{\sc i} column
density shows the emission to be an arc-like structure, with the 
highest column density gas towards the star forming regions (Fig. 5). 
\cite{Laurikainen} identify two regions, one in the north and the other in the south of the disk of NGC7753, as regions of high star formation. 
The Spitzer images \citep{Smith} are consistent with the regions identified by 
\cite{Laurikainen}. The H{\sc i} column density in these regions are amongst the
highest and the peaks range from $\sim$1$-$4 $\times$10$^{21}$ cm$^{-2}$.  
The galaxy NGC7752 and the small galaxy 2MASX J23470758+2926531, in the eastern side also appears bright in the Spitzer image, 
possibly due to star formation induced by the interaction. 

Of all the Spitzer bands, the 24 micron band seems to be the best suited for tracing recent star formation. This band is dominated by emission from very small grains heated by the UV radiation field. \cite{Calzetti} find a tight correlation between Pa$\alpha$ and 24 micron flux density, implying 24 micron flux density is well suited for tracing the current H{\sc ii} regions. The 8 micron flux density, which is commonly associated with the larger sized PAHs, also correlates with the Pa$\alpha$ flux density, but with some non-linearity. To trace SF regions, this band may not be the best as the large PAH molecules are often destroyed by the high intensity ionising radiation deep inside an H{\sc ii} region. Using the empirical formula from \cite{Calzetti} and the 24 micron flux densities from \cite{Smith}, we derive the SFR in Arp86. \cite{Smith} provides the global 24 micron flux density from NGC7753, NGC7752 and the tidal bridge. Accordingly the SFRs associated with NGC7753, NGC7752 and the tidal bridge are 9.0, 7.1 and 0.6 M$_{\odot}$ yr$^{-1}$ respectively. 

\subsection{Threshold H{\sc i} column densities}

\cite{Kennicutt} showed that for a self gravitating gas disk, there must be a threshold 
H{\sc i} column density for star formation to set in. The range of this threshold density varies in the range 10$^{20}$ cm$^{-2}$ $-$ 10$^{21}$ cm$^{-2}$ \citep{Kennicutt}. 
For irregular galaxies, \cite{Skillman} found this limit to be 10$^{21}$ cm$^{-2}$, 
averaged over 500 pc within the disks. They also found that for giant H{\sc i} 
regions to form, the gas density needs to be few times more than this threshold 
density and below this threshold star formation is actually suppressed in the 
observed galaxies. Recent observations over the last decade contain evidence of star 
formation happening in a varied range of densities. The H{\sc i} bridge in the Magellanic Clouds, which has typical column densities $\sim$10$^{20}$ $-$10$^{21}$ cm$^{-2}$ is 
known to have star formation happening in it \citep{Harris}. However, no intense 
star formation is seen to happen in the more diffuse Magellanic stream with N(H{\sc i}) 
$\le$ 3$-$5 $\times10^{20}$ cm$^{-2}$ \citep{Bruns}. In denser environments, 
\cite{Hibbard} reported intense star formation in the tidal bridges of the Antennae. 
\cite{Maybhate} report of a critical H{\sc i}  column density value of 
N(H{\sc i})$>$ 4$\times$10$^{20}$ cm$^{-2}$ over kiloparsec scale for super star clusters to form. More recently \cite{de Mello} reported star formation in the tidal bridge between
 M81 and M82, where N(H{\sc i}) is $\sim$5$-$30$\times 10^{20}$ cm$^{-2}$. All these 
observations more or less agree with the fact that high H{\sc i} column density regions 
are associated with the high star formation zones in the disks of galaxies. The thresholds found in different studies could be different for two reasons. 
Firstly, the scales over which the average H{\sc i} column densities are quoted are not the
same for all observations. Bigger areas can dilute the column densities and therefore the numbers quoted will reflect a lower threshold than the actual one. And secondly, the 
threshold density \citep{Toomre} may be a crucial parameter for forming stars, but it may not be the sole defining factor for star formation to set in \citep{Kennicutt}.

Our observations of star formation happening in the Arp86 system is in general agreement 
with the above-mentioned observations. 
We do find that star forming zones are associated with regions of high density H{\sc i} clouds in the disks of the two galaxies of Arp86. Even in the tidal bridge (SFR $\sim$0.6 M$_{\odot}$ yr$^{-1}$); which is a region of N(HI) $\sim$1$-$1.5 $\times$10$^{21}$ cm$^{-2}$  averaged over scales of $\sim$3 kpc, we note a positive correlation of the Spitzer bright regions to high H{\sc i} column density regions.

\subsection{H{\sc i} velocity dispersion}

NGC7753 and NGC7752 have inclination angles $i \sim$49$^\circ$ and 75$^\circ$ respectively.
Fig. 4 shows the dispersion in NGC7752 to be high ($>$ 60 km s$^{-1}$), 
possibly due to the inadequate resolution of the observations. 
In NGC7753, the velocity dispersion ranges from 6 km s$^{-1} -$50 km s$^{-1}$. 

The two regions of intense star formation identified by \cite{Laurikainen}, have a dispersion $\le$ 25 km s$^{-1}$. But some areas in the tidal bridge, where star formation is going on, has a higher value of dispersion $\sim$50 km s$^{-1}$. 
Opinion, about whether dispersion should be high or low in the star forming zones, is 
divided in literature. Though there are some suggestions that dispersion should 
be low for the gas to gravitationally collapse and form stars, observational evidences 
point towards a varied range of dispersion values in the star forming zones. While some 
observations have shown high H{\sc i} dispersion to correlate with regions of low 
or no star formation \citep{Ryan-Weber}, some have seen enhanced dispersion 
($\sim$40 km s$^{-1}$) to correlate with H{\sc ii} regions \citep{Irwin}. Also a 
simple correlation of dispersion and star forming region should not be expected as 
it is difficult to identify the cause-effect relation between the two. 
In principle high dispersion can lead to high star formation as the Jeans mass 
is proportional to the fouth power of dispersion \citep{Elmegreen} or it can be 
an effect of high star formation, if the cloud is randomized by the outflows of massive star formation activity \citep{Irwin}. The situation in this system is also unclear.
We find regions of low dispersion in the disk of NGC7753, medium dispersion 
in the tidal bridge and high dispersion in NGC7752 and the galaxy 
2MASX J23470758+2926531, and all of these being star forming regions. Also,
we see high velocity dispersion of $>$ 60 km s$^{-1}$ towards the eastern
edge of the south-eastern feature, without any evidence of star formation.

\begin{figure}
    \hspace{.1cm}
    \centerline{\rotatebox{-90}{\includegraphics*[height=3.5in]{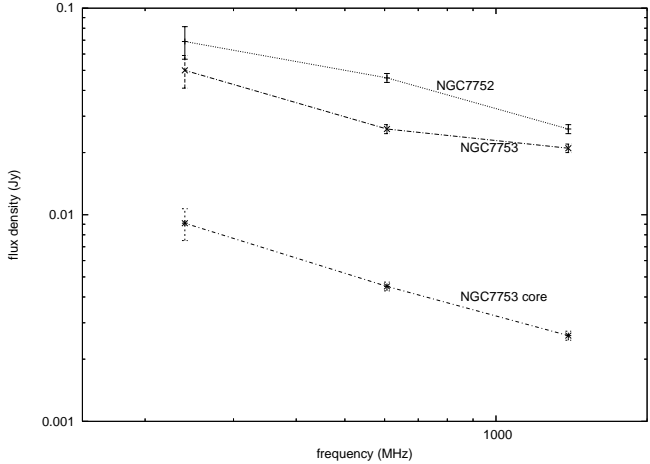}} }
    \vspace*{0.05in}
    \caption{Continuum spectra from the Arp86 system.}
    \label{}
    \end{figure}

\subsection{2MASX J23470758+2926531: a tidal dwarf galaxy?}

As discussed in Section 3, our observations show the galaxy 2MASX J23470758+2926531
to be a part of the Arp86 system.
We estimate the stellar mass of this galaxy using its K$-$band magnitude and 
K$-$J colours, listed in NED. The mass to light ratio of galaxies in the K band 
($M/L_{K}$) are related to the K$-$J colours by the equation \citep{bell}

\begin{equation} 
log(M/L_{K})= a_{K}+ b_{K} ~colour_{K-J}, 
\end{equation}

while the luminosity in the K-band, $L_{K}$ is related to the absolute magnitude 
in the K-band ($M_{K}$) as \citep{worthey}

\begin{equation}
L_{K}(L_{\odot})= \exp {(0.921034~ (3.33 - M_{K}))}.
\end{equation}

\begin{figure}
    \hspace{.1cm}
    \centerline{\rotatebox{-90}{\includegraphics*[height=3.5in]{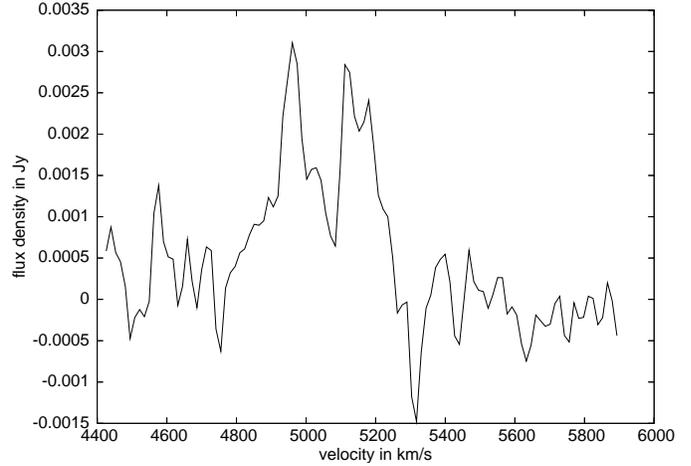}} }
    \vspace*{0.05in}
    \caption{H{\sc i} spectrum of 2MASX J23470758+2926531}
    \label{}
    \end{figure}

The estimated absolute K magnitude is $-$20.3 and the stellar mass of this galaxy 
is 2.9$\times$10$^{9}$ M$_{\odot}$. The H{\sc i} mass, estimated from the 
spectrum (Fig. 9) is 4.5$\times$10$^{8}$ M$_{\odot}$. 
The spectrum suggests a weak double horn, consistent with rotation, although the
spectrum could be affected by more extended tidal features near the galaxy.  
It is interesting to enquire whether this galaxy might be a tidal dwarf galaxy (TDG).
TDGs are known to be comparable in luminosity to typical dwarfs, though metallicities 
can be higher in TDGs. They have characteristic blue colours as a result of active 
starburst, high velocity dispersion and can have high H{\sc i} masses 
$\sim$ few times 10$^{8}$ M$_{\odot}$ and stellar masses $\sim$ few times 
10$^{9}$ M$_{\odot}$ \citep{Duc}. Our TDG candidate is similar to the above stated 
galaxies in terms of stellar and H{\sc i} mass, physical extent and high dispersion 
values ($\ge$ 50 km s$^{-1}$). The galaxy is bright in the Spitzer and Galex images, 
indicating recent star formation. The properties are consistent with that of a TDG.   
The dynamical mass of 2MASX J23470758+2926531 has been estimated using the rotation
velocity as observed in the H{\sc i} spectrum and a radial extent $\sim$ 12.5$^{\prime\prime}$.
This is half the beam width of our H{\sc i} image (Fig. 4) and corresonds to $\sim$4 kpc.
The mass thus derived is 1.4$\times$10$^{10}$ M$_{\odot}$. This makes the 
M$_{dynamical}$/L$_{Kband}$ ratio to be 4.8. For TDG candidates this number is expected 
to be small due to lack of dark matter. The median value for the mass to light ratios of
the tidal dwarf galaxy candidates in Stephan's Quintet was found to be 7.0 \citep{amram}. Judging by
these arguments it is likely that 2MASX J23470758+2926531 is a candidate TDG in 
Arp86 system. However, occurrence of TDGs in such unequal mass systems is very rare. Most of the 
known candidate TDGs have formed during roughly equal mass strong interactions which has brought 
out a large amount of gas out of the galaxies' disks, e.g. TDGs in Arp105 \citep{Ferreiro}, NGC3227/3226 \citep{Mundell},
Arp245 \citep{Duc}. Garland, formed in the interaction of NGC3077/M81 is one of the few known TDG
candidates to form in an encounter involving highly unequal mass galaxies \citep{garland}. 
Although we suggest this to be a TDG candidate from the available information, the  
possibility that this might be a small galaxy which has been part of the Arp86 group cannot be ruled out.

\subsection{Radio continuum emission}
As mentioned earlier, radio continuum emission has been detected from both NGC7752 and
NGC7753, and from a bridge between the two galaxies, but no emssion has been detected
from the candidate TDG. 
The spectrum of radio emission from NGC7752 shows some evidence of steepening towards higher
frequencies.  The spectral indices (SI) of NGC7752 is $\sim$0.4 between 240 and 606 MHz 
and $\sim$0.7 between 606 and 1394 MHz. This is consistent with the 
finding of \cite{hummel} who estimates a median break frequency of approximately 700 MHz
for a sample of spiral galaxies. 
A shallow low-frequency spectral index of $\sim$0.4 for the disk emission in NGC7752 is on the lower side.
In the sample of 27 spiral galaxies studied by \citep{hummel}, only three of his
galaxies have a low-frequency (below $\sim$1 GHz) SI less than 0.4. 
Modelling the steepening of the spectrum in the sample of sources as being due to
propagation and energy loss processes, the author equates the asymptotic low-frequency
SI to the injection spectral index and finds the values to be in the range of 0.3 to 0.6
with a median value of $\sim$0.40$\pm$0.05. Our low-frequency SI of NGC7752 is consistent with
this interpretation. 

The SI of the extended emission 
in NGC7753 is $\sim$0.7 between 240 and 606 MHz, similar to the low-frequency SIs
of several of the galaxies studied by \citep{hummel}. 
However, the radio emission from this galaxy shows evidence
of flattening between 606 and 1394 MHz, with a SI of 0.25, which needs to be confirmed from higher-frequency
observations. The 1400-MHz flux density from NVSS is also consistent with this flattening
of the higher frequency radio spectrum. This would suggest that thermal free-free emission 
could also play a significant role in determining the spectral shape. This is also suggested by the 
rather high SFR of 9.0 M$_{\odot}$ yr$^{-1}$ in this galaxy. It is relevant to note that the galaxies in the 
sample of \cite{hummel} which show evidence of flattening of the high-frequency spectrum 
are NGC253, NGC1569 and NGC5236, all of which show evidence of a strong starburst.

The radio 1400-MHz and far-infrared (FIR) 60 $\mu$m luminosities of NGC7752 are 22.1 and 10.3 (in log scale) respectively,
while the corresponding values for NGC7753 are 20.3 and 8.3 respectively. These values are
consistent with the radio-FIR correlation \citep{yun}. This is an expected result since none of these galaxies have prominent active galactic nuclei. The radio luminosities at 1400 MHz 
indicate a supernova rate of 0.10 and 0.08 for NGC7752 and NGC7753 respectively using the 
formalism of \cite{yin}. These numbers are consistent with the supernova rate estimates made using lower frequency flux densities as well. The supernova rates for NGC7752 and NGC7753 appear similar to estimates for
other galaxies with different degrees of starburst activity. These estimates
include $\sim$0.04 to 0.1 yr$^{-1}$ for a small sample of galaxies including the
archetypal starburst galaxy NGC1808 \citep{Collison},
$\sim$0.1 yr$^{-1}$ for M82 \citep{huang}, $\le$0.1 to 0.3 for
NGC253 \citep{ulv}, $\sim$0.1 yr$^{-1}$ for the irregular
starburst galaxy Mrk 325 \citep{yin} and the starburst galaxy
NGC3448 of
the Arp 205 system \citep{nor}, $\sim$0.07 yr$^{-1}$ for
NGC6951
\citep{saikia} and $\sim$0.14 yr$^{-1}$ for the superwind galaxy
NGC1482 \citep{hota}. 

For both NGC7752 and NGC7753 we have estimated the minimum energy, equipartition magnetic field and spectral
ages by integrating the spectrum between 10 MHz and 100 GHz, a filling factor of unity
and an electron to proton energy density ratio of unity. For NGC7753, whose high-frequency 
spectrum is likely to be affected by thermal emission, we have assumed a SI of 0.7 between
10 MHz and 100 GHz. The minimum energy and the equipartition magnetic fields for NGC7752 are 8.1 $\times$ 10$^{53}$ ergs and 3.6 $\mu$gauss and for NGC7753 are 5.9 $\times$ 10$^{54}$ ergs and 2.6 $\mu$gauss. The spectral ages of NGC7752 and NGC7753 for a break frequency of 600 MHz are $\sim$1.1 $\times$ 10$^{8}$ years and 1.7 $\times$ 10$^{8}$ years respectively.

The radio continuum emission also shows a bridge of emission between the two galaxies
which is seen clearly in the 606-MHz image with a resolution of 45 arcsec.
Although H{\sc i} bridges have been seen in many interacting systems, radio continuum 
bridges are less common. This could be at least partly due to the absence of a
systematic search for radio continuum bridges at low frequencies, where these bridges
are expected to be more prominent due to their steep spectral indices \citep{Condon,Condon3,Kantharia}.  
Also for the bridge to be bright in synchrotron emission, either in-situ star formation 
is necessary or the tidally disrupted bridge material should have undergone a recent phase 
of star formation, to give rise to the synchrotron emitting particles. One of the early 
systems with a radio continuum bridge was
Ho 124 \citep{van der Hulst}. Since then, several other interacting galaxies have 
been detected with radio bridges such as the Taffy galaxies \citep{Condon}. Different
processes have been suggested to explain the radio continuum bridges. \cite{Condon} 
have suggested that the bridge magnetic fields and relativistic particles have been
stripped from the interpenetrating disks during a nearly head-on collision in the Taffy
galaxies. For Ho 124, \cite{Kantharia} have suggested that the bridge is likely to be
of tidal origin with no evidence of significant star formation in the bridge. In 
the case of Arp86, both the H$\alpha$ \citep{marcelin} and mid-infrared \citep{Smith}
images show evidence of star formation in the bridge region. This suggests that the radio
continuum bridge could have a significant contribution from star formation between the two galaxies. As mentioned
earlier, the star-formation rate in the bridge is 0.6 M$_\odot$ yr$^{-1}$. Deeper images
at the other frequencies are required to estimate the spectral index of the bridge.  

In the highest resolution map (7$^{\prime\prime} \times$ 3$^{\prime\prime}$) at 
1394 MHz the core of NGC7753 is a resolved source. The 
spectral index estimated using the 16$^{\prime\prime}$ images, which is the
resolution of the 240-MHz image, is $\sim$0.7. The detection of a radio
source at the centre of the galaxy is consistent with the the identification of
a LINER nucleus, a mildly active galaxy. Although the spectral index is steep,
unlike the cores of powerful radio galaxies and quasars, steep spectral indices
have been seen in the cores of several mildly active galaxies such as the 
Seyfert galaxies \citep{Kukula,thean}. This is possibly due to the contamination of the
flat-spectrum nucleus by more extended emission.   

\section{Conclusions} 
We present the results of the GMRT observations 
of the interacting system Arp86 in both neutral atomic hydrogen, H{\sc i}, and in radio 
continuum at 240, 606 and 1394 MHz.  H{\sc i} maps of the system reveal disturbed morphology with extended tidal arms and debris suggesting an M51-like interaction. The star forming regions in the system seem to correlate well with regions of high H{\sc i} column density. We detect a possible TDG candidate in the system. The tidal bridge is also detected in radio continuum emission. The system is undergoing intense star formation and this probably reflects in an unusually flat SI for the disk emission from NGC7753 between 606 and 1394 MHz. However, this flattening needs to be confirmed with higher frequency observations.

\section{Acknowledgments}
\noindent We thank the reviewer for his/her useful comments and suggestions. We thank the staff of the GMRT who have made these observations possible. The GMRT is operated by the National Centre for Radio Astrophysics of the Tata Institute of Fundamental Research. This research has made use of the NASA/IPAC Extragalactic Database (NED) which is operated by the Jet Propulsion Laboratory, California Institute of Technology, under contract with the National Aeronautics and Space Administration. CS would like to thank Jayaram Chengalur for some very useful discussions.

\end{document}